\documentstyle[12pt]{article}
\def\beq{\begin{equation}}
\def\eeq{\end{equation}}

\begin{document}

\begin{titlepage}

\begin{center}
{\Large\bf   Neutrino Mixings and Fermion Masses \\
in Supersymmetric SU(5)
}
\end{center}
\vspace{0.5cm}
\begin{center}
{\large Qaisar Shafi$^{a}$\footnote {E-mail address:
shafi@bartol.udel.edu} {}~and
{}~Zurab Tavartkiladze$^{b}$\footnote {E-mail address:
z\_tavart@osgf.ge} }
\vspace{0.5cm}

$^a${\em Bartol Research Institute, University of Delaware,
Newark, DE 19716, USA \\

$^b$ Institute of Physics, Georgian Academy of Sciences,
380077 Tbilisi, Georgia}\\
\end{center}

\vspace{1.0cm}

\begin{abstract}
We consider neutrino mixings in supersymmetric $SU(5)$ 
supplemented by a ${\cal U}(1)$ flavor symmetry.
In particular, we show how bi-maximal neutrino mixings
can be realized. Two scenarios for implementing the small mixing angle
MSW solution, 
with one involving a sterile state $\nu_s$ and maximal
$\nu_{\mu }-\nu_s$ mixing to
resolve the atmospheric anomaly, are also discussed.
Finally, a new mechanism for eliminating the asymptotic relations
$m^{(0)}_{\mu}\ =m^{(0)}_s$ and $m^{(0)}_e=m^{(0)}_d$, 
while retaining $m^{(0)}_{\tau }=m^{(0)}_b$, is presented.
It employs `matter' multiplets in the
$15+\overline{15}$ of $SU(5)$, and is consistent with the various
oscillation scenarios, as well as with the unification at $M_{GUT}$
of the three gauge couplings.

\end{abstract}

\end{titlepage}


%

The recent atmospheric neutrino results \cite{atmSK} from Superkamiokande
(SK)
have
greatly enhanced the case for new physics beyond the standard model (SM).
The data is found to be consistent with a two-flavor model involving
$\nu_{\mu}-\nu_x$ oscillations, where
$\nu_x$ could either be
$\nu_{\tau}$ or a (new) sterile neutrino
$\nu_s$.  The mixing angle is estimated to be large,
$sin^22\theta\stackrel{_>}{_\sim} 0.8$, with
$\Delta m^2\sim10^{-2}-10^{-3}$eV$^2$.  It is difficult to see how neutrino
masses of order
$10^{-1}$eV (or larger, in case some masses are degenerate) can arise from
the SM or its minimal supersymmetric extension (MSSM), without involving new
physics at some `intermediate' mass scale.  Non-zero neutrino masses from
Planck scale
($M_{P}=2.4\times10^{18}$~GeV) suppressed dimension five operators are
expected to be 
$\sim10^{-5}$eV (or so).

The solar neutrino data offers independent evidence for neutrino
oscillations \cite{solSK}.
The data is consistent with either vacuum oscillations in which
$sin^22\theta\sim0.7-1.0$ and $\Delta m^2\sim5\times10^{-11}-10^{-10}$eV$^2$,
or the small mixing angle MSW solution with
$sin^22\theta\sim({\rm few})\cdot10^{-3}$ and
$\Delta m^2\sim({\rm few})\cdot10^{-6}$eV$^2$. It is found that 
the data favors a vacuum
solution involving only active neutrinos, while the
MSW solution can involve a sterile neutrino $(\nu_e\rightarrow\nu_s)$. 

In a recent paper \cite{numssm}, we presented a systematic approach,
based on
$\nu$MSSM
(MSSM augmented with the seesaw mechanism) and its extension for incorporating
the various two-flavor neutrino oscillations allowed by the atmospheric and
solar neutrino data.  It utilizes in an essential way a flavor
${\cal{U}}(1)$ symmetry \cite{u1}, whose spontaneous breaking by (the
scalar
component of) a MSSM singlet field X yields an important `expansion' parameter
$\epsilon\equiv\langle X\rangle/M_P(\simeq0.2)$.  The
${\cal U}(1)$ symmetry together with
$\epsilon$, also plays an important role in understanding the quark and
charged lepton mass hierarchies, as well as the magnitudes of the CKM matrix
elements.  The approach we followed was especially guided by the desire to
realize maximal mixing
($sin^22\theta=1$) to explain the atmospheric anomaly, which seems favored
by the SK data.  A general mechanism for achieving this was presented
in ref. \cite{numssm}, and it was noted that except for
$\nu$MSSM and its SU(5) extension, this typically required the existence
of a new (sterile) neutrino in order to simultaneously account for the solar
and atmospheric neutrino puzzles.  For instance, maximal
$\nu_{\mu}-\nu_{\tau}$ oscillations can be accompanied by the small mixing
angle MSW solution involving
$\nu_e$ and $\nu_s$.  It was pointed out that a
${\cal U}(1)-{\cal R}$ symmetry can be particularly effective at keeping a
sterile neutrino light.

 The purpose of this letter is twofold. First, we would like to
demonstrate
how a variety of two-flavor neutrino oscillations scenarios,
that are compatible with the current atmospheric and solar neutrino
data, can be
realized within a $SU(5)$ setting, supplemented by singlet
(right handed) superfields and a ${\cal U}(1)$ flavor symmetry. In
particular,
we show how bi-maximal mixings \cite{bimax} among the active 
neutrinos, which is consistent with the latest atmospheric and
solar neutrino experiments \cite{bar}, can be realized.  To do this, we 
first utilize a version
of the seesaw mechanism discussed in \cite{sk} to obtain large
(including maximal)
mixing in the $\nu_{\mu }-\nu_{\tau }$ sector, with only one neutrino
acquiring a non-zero mass. We then invoke the mechanism described in
\cite{numssm} to implement
maximal $\nu_e-\nu_{\mu , \tau }$ mixing to resolve the solar neutrino
puzzle. This is followed by a discussion of how the small mixing 
angle MSW solution can be
realized in two
distinct ways, with one (non-minimal) approach yielding a light
sterile state and gives possibility of existence of the neutrino 
hot dark matter.
Our second goal here is to provide a new mechanism for eliminating
the unacceptable
asymptotic relations predicted in $SU(5)$ with minimal higgs,
to wit, $m_{\mu}^{(0)}=m_s^{(0)}$ and $m_e^{(0)}=m_d^{(0)}$,
retaining in the process the desirable relation
$m_{\tau }^{(0)}=m_b^{(0)}$. This we
achieve by introducing a pair of $15 + \overline{15}$  `matter'
multiplets (rather than a $45$-plet of higgs \cite{su52}
as is commonly done).
It should be stressed that this mechanism is consistent with the
various oscillations scenarios, and
it also preserves the unification at $M_{GUT}$ of the three gauge
couplings.

The minimal SU(5) model contains the following `matter' multiplets 
(we assume a $Z_2$ `matter' parity which distinguishes the 
`matter' and
`higgs' supermultiplets):

\begin{equation}
10_{\alpha }=(q,~u^c,~e^c)_{\alpha }~,~~~~~
\bar 5_{\alpha }=(d^c,~l)_{\alpha }~,
\label{reps}
\end{equation}
where $\alpha =1, 2, 3$ denotes the flavor index. The transformation
properties
under $SU(3)_c\times SU(2)_L$  of the various superfields are:

\begin{equation}
q(3,~2)~,~u^c(\bar 3,~1)~,~~e^c(1,~1)~,~~d^c(\bar 3,~1)~,~l(1,~2)~.
\label{trans}
\end{equation}

The higgs multiplets consist of
$\Sigma(24),~H(5),~\bar{H}(\bar 5)$, where the 24-plet breaks
$SU(5)$ to $SU(3)_c \times SU(2)_L \times U(1)_Y$,
and
$H(5)$ $(\bar H(\bar 5))$
contains the electroweak doublet $h_u$ ($h_d$) which provides masses for
the up
quarks (down quarks and charged leptons).

We now introduce a
${\cal{U}}(1)$ flavor symmetry with a judicious choice of flavor
charges
which yield the observed quark and charged lepton mass
hierarchies, as well as the magnitudes of the CKM matrix elements.
For the asymptotic Yukawa couplings we will assume the following ratios:

$$
\lambda_u : \lambda_c :  \lambda_t \sim
\epsilon^6 : \epsilon^4 :1~,
$$
$$
\lambda_d :\lambda_s :\lambda_b \sim
\epsilon^5:\epsilon^2 :1~,
$$
\begin{equation}
\lambda_e :\lambda_{\mu } :\lambda_{\tau } \sim
\epsilon^5:\epsilon^2 :1~.
\label{lambdas}
\end{equation}
For the CKM matrix elements we take
\beq
V_{us} \sim \epsilon \ , \ V_{cb} \sim \epsilon^2 \ ,\ V_{ub} \sim \epsilon^3
\label{ckm}
\eeq
One of our goals is to achieve large (including maximal) mixing in the
$\nu_{\mu } -\nu_{\tau}$ sector \cite{ram}.

These considerations determine the following Yukawa matrix structure
for the down quarks and charged leptons (the matrix elements are determined
only up to factors of order unity; for simplicity, we will ignore 
CP violation):

\begin{equation}
\begin{array}{ccc}
 & {\begin{array}{ccc}
\hspace{-5mm}~10_1& \,\,10_2 & \,\,10_3

\end{array}}\\ \vspace{2mm}
{Y}_{d,e}\sim \begin{array}{c}
\bar 5_1 \\ \bar 5_2 \\ \bar 5_3
 \end{array}\!\!\!\!\! &{\left(\begin{array}{ccc}
\,\,\epsilon^5 ~~ &\,\,\epsilon^4~~ &
\,\,\epsilon^2
\\
\,\,\epsilon^3 ~~  &\,\,\epsilon^2~~ &
\,\,1
 \\
\,\,\epsilon^3~~ &\,\,\epsilon^2~~ &\,\,1
\end{array}\right)\epsilon^a }~,
\end{array}  \!\!  ~~~~~
\label{lepdown}
\end{equation}
where $a=0,1,2$ determines the value of the MSSM parameter
$\tan \beta $($\sim \frac{m_t}{m_b}\epsilon^a$).
The texture in (\ref{lepdown}) is realized for the following
prescription of
the ${\cal U}(1)$ charges of the various supermultiplets:

$$
Q_{\bar 5_3}=Q_{\bar 5_2}=q_1~,~~~~~Q_{\bar 5_1}=q_1-2~,
$$
$$
Q_{10_3}=q_2~,~~~Q_{10_2}=q_2-2~,~~~
Q_{10_1}=q_2-3~,
$$
\beq
Q_{\bar H}=-a-q_1-q_2~,~~~Q_{H}=-2q_2~.
\label{charges}
\eeq
Here and below the ${\cal U}$(1) charges of the 
supermultiplets are presented in units of the
charge of the $X$ superfield ($Q_X=1$). The charges $q_1$ and $q_2$ are 
arbitrary
for the time being. Note that
$\bar{5}_2$ and $\bar{5}_3$ carry the same
${\cal{U}} (1)$ charge in order to accommodate large mixing in the
$\nu_{\mu}-\nu_{\tau}$ sector \cite{ram}.

Similarly, taking into account (\ref{charges}), the up quark Yukawa matrix  
has the following texture

\begin{equation}
\begin{array}{ccc}
 & {\begin{array}{ccc}
\hspace{-5mm}~~10_1 & \,\,10_2  & \,\,10_3

\end{array}}\\ \vspace{2mm}
{Y}_u\sim \begin{array}{c}
10_1 \\ 10_2 \\10_3
 \end{array}\!\!\!\!\! &{\left(\begin{array}{ccc}
\,\,\epsilon^6~~ &\,\,\epsilon^5~~ &
\,\,\epsilon^3
\\
\,\,\epsilon^5~~   &\,\,\epsilon^4~~  &
\,\,\epsilon^2
 \\
\,\,\epsilon^3~~ &\,\,\epsilon^2~~ &\,\,1
\end{array}\right) }~.
\end{array}  \!\!  ~~~~~
\label{up}
\end{equation}

We see that large (hopefully even maximal) mixing can arise in 
the $\nu_{\mu}-\nu_{\tau}$ sector
in a rather straightforward manner \cite{ram}.  
In order to achieve the desired mass splitting
$\Delta m^2 \sim 10^{-2}-10^{-3}$ eV$^2$,
let us choose $q_1-2q_2=0$, and introduce a SU(5) singlet superfield N,
with $Q_N=0$.
Consider the superpotential couplings:

\beq
W_N= M_NN^2+(a \epsilon^2\bar 5_1+b\bar 5_2+c\bar 5_3)H N\ ,
\label{N}
\eeq
where the coefficients a,b,c are all of order unity.
Through the seesaw
mechanism we obtain the following mass for the `light' mass eigenstate 
\beq
m_{\nu_{3}}\sim \frac{h^2_u}{M_N}
\sim \sqrt{\Delta m^2_{\mu \tau}}.
\label{mas3}
\eeq
Clearly, the two remaining states are still massless.

In order to achieve large (especially maximal) mixing, say  between
$\nu_e-\nu_{\mu}$, to resolve the solar neutrino puzzle, we could try
assigning a zero ${\cal{U}} (1)$ charge to
$\bar{5}_1$.  However, coupled with the charge assignment $10_1(-5)$, 
this implies
$V_{us}\sim\epsilon^3$ which is unacceptable! (In
$\nu$MSSM this strategy may work because the quarks and leptons 
belong to separate multiplets).
We are thus led to exploit the maximal mixing scenarios discussed
in ref \cite{numssm,422}.
We introduce two new SU(5) singlet states
${\cal N}_1, {\cal N}_2$ with charges $Q_{{\cal N}_1}=-2$ and
$Q_{{\cal N}_2}=2$,
and consider the following two (Dirac and Majorana)
matrices:

\begin{equation}
\begin{array}{cc}
 & {\begin{array}{cc}
{\cal N}_1~&\,\,{\cal N}_2~~~~~~
\end{array}}\\ \vspace{2mm}
\begin{array}{c}
\bar 5_1\\ \bar 5_2 \\ \bar 5_3

\end{array}\!\!\!\!\! &{\left(\begin{array}{ccc}
\,\, \epsilon^4~~ &
\,\,  1
\\
\,\, \epsilon^2~~ &\,\,0
\\
\,\, \epsilon^2~~ &\,\,0
\end{array}\right)\kappa H }~,
\end{array}  \!\!~~~
\begin{array}{cc}
 & {\begin{array}{cc}
{\cal N}_1~&\,\,
{\cal N}_2~~~~~
\end{array}}\\ \vspace{2mm}
\begin{array}{c}
{\cal N}_1 \\ {\cal N}_2

\end{array}\!\!\!\!\! &{\left(\begin{array}{ccc}
\,\, \epsilon^4
 &\,\,~~~1
\\
\,\, 1
&\,\,~~~0
\end{array}\right)M_{\cal N}~,
}
\end{array}~~~
\label{Ns}
\end{equation}
where $\kappa $ is some dimensionless coefficient, and
$M_{\cal N}$ is an appropriate heavy scale.
In the basis in which the couplings (\ref{lepdown}) (responsible for
generation of the charged lepton masses) are diagonal, the
first matrix in (\ref{Ns}) will be  modified, and the appropriate
`Dirac' and `Majorana' couplings
will have the forms:

\begin{equation}
\begin{array}{cc}
m_D=\!\!\!\!\! &{\left(\begin{array}{ccc}
\,\, \epsilon^4~ &
\,\,  1
\\
\,\, \epsilon^2~ &\,\,\epsilon^2
\\
\,\, \epsilon^2~ &\,\,\epsilon^2
\end{array}\right)\kappa h_u }~,
\end{array}
\begin{array}{cc}
%
~~M_R=\!\!\!\!\! &{\left(\begin{array}{ccc}
\,\, \epsilon^4
 &\,\,1
\\
\,\, 1
&\,\,0
\end{array}\right)M_{\cal N}~.
}
\end{array}
\label{dirmaj}
\end{equation}
Note that the hierarchical structure of the couplings in (\ref{N}) 
is unchanged.
Working in this basis, the matrix which diagonalizes the `light' neutrino mass matrix
will coincide with the physical lepton mixing matrix.

Together with
(\ref{N}) and (\ref{Ns}), the  ${\cal U}(1)$ symmetry also permits 
the coupling
$M'\epsilon^2{\cal N}_1N$. Assuming

\begin{equation}
M_N\gg M'\epsilon^2~,~~~~~
M_{\cal N}\stackrel{_>}{_\sim } M'^2/M_N~,
\label{cond}
\end{equation}
after integrating out the $N$ and
${\cal N}_{1,2}$ states, the seesaw mechanism yields:

\beq
\hat{m}_{\nu }=\hat{A}m+\hat{B}m'~,
\label{matnu}
\eeq
where

\beq
m\equiv \frac{h_u^2}{M_N}~,~~~~
m'\equiv\frac{\kappa^2\epsilon^2h_u^2}{M_{\cal N}}~,
\label{scales}
\eeq
and

$$
\begin{array}{ccc}
 & {\begin{array}{ccc}
~& \,\,~  & \,\,~~
\end{array}}\\ \vspace{2mm}
\hat{A}=
\begin{array}{c}
\\  \\
 \end{array}\!\!\!\!\!\!\!\!&{\left(\begin{array}{ccc}
\,\,a^2\epsilon^4  &\,\,~~ab\epsilon^2 &
\,\,~~ac\epsilon^2
\\
\,\,ab\epsilon^2   &\,\,~~b^2  &
\,\,~~bc
 \\
\,\, ac\epsilon^2 &\,\,~~bc  &\,\,~~c^2
\end{array}\right)m }~,
\end{array}  \!\!  ~~
$$
\beq
\begin{array}{ccc}
 & {\begin{array}{ccc}
~~\,\,~  & \,\,~~~~
\end{array}}\\ \vspace{2mm}
\hat{B}=
\begin{array}{c}
\\  \\
 \end{array}\!\!\!\!\!\!\!\!&{\left(\begin{array}{ccc}
\,\,\epsilon^2 &\,\,~1 &
\,\,~1
\\
\,\,1   &\,\,~\epsilon^2  &
\,\,~\epsilon^2
 \\
\,\, 1 &\,\,~ \epsilon^2  &\,\,~\epsilon^2
\end{array}\right)m' }
\end{array}  \!\!  ~.
\label{AB}
\end{equation}
Note that ${\rm Det}\hat{B}=0$ since $\hat{B}$ is obtained through the
exchange
of the two heavy ${\cal N}_{1, 2}$ states. 
The matrix $\hat{A}$ has only one
nonzero eigenvalue.

For $\kappa \sim 1$,
$M_{\cal N}\sim 4\cdot 10^{16}$~GeV and $M_N\sim 10^{15}$~GeV
(the conditions in (\ref{cond}) are satisfied
for $M'< 10^{15}$~GeV) one obtains

$$
m_{\nu_3}\simeq m(b^2+c^2+a^2\epsilon^4)
\sim 3\cdot 10^{-2}~{\rm eV}~,
$$
\beq
m_{\nu_1 }\simeq m_{\nu_2 }\simeq m'\sim 3\cdot 10^{-5}~{\rm eV}~.
\label{masses}
\eeq

Ignoring CP violation the neutrino mass matrix (\ref{matnu})
can be diagonalized by the
orthogonal transformations $\nu_{\alpha }=U_{\nu}^{\alpha i}\nu_i$, where
$\alpha =e, \mu, \tau $ denotes flavor indices, and $i=1, 2, 3$
the mass eigenstates.
One finds 

\beq
\begin{array}{ccc}
U_{\nu }=~~
\!\!\!\!\!\!\!\!&{\left(\begin{array}{ccc}
\,\,\frac{1}{\sqrt{2}} &\,\,~~\frac{1}{\sqrt{2}} &
\,\,~~s_1
\\
\,\,-\frac{1}{\sqrt{2}}c_{\theta }  &\,\,~~~~\frac{1}{\sqrt{2}}c_{\theta }
&
\,\,~~s_{\theta }
 \\
\,\, ~~\frac{1}{\sqrt{2}}s_{\theta } &\,\,~
-\frac{1}{\sqrt{2}}s_{\theta }  &\,\,~~c_{\theta }
\end{array}\right) }
\end{array} +~{\cal O}(\epsilon^2)~,
\label{lepckm}
\end{equation}
with

\beq
\tan \theta =\frac{b}{c}~,~~~~s_1=\frac{a\epsilon^2}{\sqrt{b^2+c^2}}~,
\label{angles}
\eeq
and $s_{\theta }\equiv \sin \theta $, $c_{\theta }\equiv \cos \theta $.

%
%
%

For the solar and atmospheric neutrino oscillation parameters we 
find

$$
\Delta m^2_{21 }\sim 2m'^2\epsilon^2\simeq 10^{-10}~{\rm eV}^2~,
$$
\beq
{\cal A}(\nu_e \to \nu_{\mu , \tau }) =1-{\cal O}(\epsilon^4)~,
\label{solosc}
\eeq
and

$$
\Delta m^2_{32}\simeq m_{\nu_3}^2\sim 10^{-3}~{\rm eV}^2~,
$$
\beq
{\cal A}(\nu_{\mu }\to \nu_{\tau })=\frac{4b^2c^2}{(b^2+c^2)^2}-
{\cal O}(\epsilon^4)~.
\label{atmosc}
\eeq
Here ${\cal A}$ determines the transition amplitude.
We have therefore realized maximal
$\nu_e-\nu_{\mu, \tau}$ mixing as promised! In the
$\nu_{\mu }-\nu_{\tau }$ system large mixing can be obtained quite
naturally for $b\sim a$, while maximal mixing holds for $b\simeq c$.
In this case the $\nu_e $ oscillations are $50\% $ into $\nu_{\mu }$
and $50\% $ into $\nu_{\tau }$. 


Next let us briefly discuss
other possible neutrino oscillation scenarios which can be realized
in the framework of $SU(5)$ GUT.
Without applying any particular mechanism, the charged
sector contribution to 
$\nu_e-\nu_{\mu, \tau}$ mixing is expected to be
$\theta_{e\mu }\sim \theta_{e\tau }\sim \epsilon^2 $ (see
(\ref{lepdown})),
which has the right magnitude for the small angle MSW solution of the
solar neutrino puzzle. By introducing only the $N$ right handed neutrino,
from (\ref{N}), (\ref{matnu}),
the neutrino mass matrix will be $m\hat{A}$, providing still large (or
even maximal) $\nu_{\mu }-\nu_{\tau }$ oscillations, and massless $\nu_1$,
$\nu_2$ states. To obtain the relevant mass scale for solar neutrino
oscillations, one can introduce a state $N'$ with ${\cal U}(1)$ charge
$Q_{N'}=0$ (we still work with the combination $q_1-2q_2=0$).
With couplings

\beq
W_{N'}= M_{N'}N'^2+(\epsilon^2\bar 5_1+\bar 5_2+\bar 5_3)H N' ,
\label{N1}
\eeq
and a suppressed term $M_1NN'$ ($M_1\ll M_N$), after integrating 
out the $N'$
state, and taking $M_{N'}\sim 10^{16}$~GeV, we get
$m_{\nu_2 }\sim h_u^2/M_{N'}\sim 10^{-3}$~eV, the desired
value for the solar neutrino oscillation parameter
($\Delta m_{21}^2\sim 10^{-6}{\rm eV}^2$).

This scenario, as well as the previously discussed case, can naturally
provide large $\nu_{\mu }-\nu_{\tau }$ mixings. For obtaining maximal
mixing to $1\% $ accuracy, the condition $b^2=c^2$ should hold to
$10\%$ accuracy. The 
mechanism of \cite{numssm} (which we applied above for obtaining maximal
$\nu_e-\nu_{\mu, \tau}$ mixings) does not work for
$\nu_{\mu }-\nu_{\tau}$ system, since these flavors carry the same
${\cal U}(1)$ charge. However, another scenario involving a 
sterile neutrino state,
suggesting maximal $\nu_{\mu }-\nu_s$ mixing for atmospheric
neutrino deficit, can still be realized. The solar neutrino puzzle can
be resolved in this case through the
small angle MSW oscillations.

To see this, consider the prescription of the ${\cal U}(1)$
charges (\ref{charges}), with $q_1-2q_2=11/2$. Introducing a right
handed  neutrino $N''$, and a light sterile state
$\nu_s$, with  charges $Q_{N''}=-11/2$,
$Q_{\nu_s}=-43/2$, the relevant couplings are

$$
W_{N''\nu_s}=\kappa'\left(\epsilon^2\bar 5_1+\bar 5_2+\bar 5_3\right)HN''+
\epsilon^{11}M_PN''^2
$$
\beq
\epsilon^{16}\left(\epsilon^2\bar 5_1+\bar 5_2+\bar 5_3\right)H\nu_s+
\epsilon^{43}M_P\nu_s^2~,
\label{st}
\eeq
where $\kappa' $ is a dimensionless coupling.
Integrating out $N''$, and using the notations

\beq
m=\epsilon^{16}h_u~,~~~m'=\frac{\kappa'^2h_u^2}{M_P\epsilon^{11}}~,~~~
m_{\nu_s}=M_P\epsilon^{43}~,
\label{not}
\eeq
the mass matrix for the light neutrinos is given by

\begin{equation}
\begin{array}{cccc}
 & {\begin{array}{cccc}
\hspace{-5mm}~~~\nu_e & \,\,~~~~\nu_{\mu }~~~  &
\,\,~\nu_{\tau }~~& \,\,~\nu_s
\end{array}}\\ \vspace{2mm}
m_{\nu }= \begin{array}{c}
\nu_e \\ \nu_{\mu } \\ \nu_{\tau } \\ \nu_s
 \end{array}\!\!\!\!\! &{\left(\begin{array}{cccc}
\,\,m'\epsilon^4  &\,\,~~m'\epsilon^2 &
\,\,~m'\epsilon^2 &~m\epsilon^2
\\
\,\,m'\epsilon^2  &\,\,~~m'  & \,\,~m'&m
 \\
\,\,m'\epsilon^2 &\,\,~~m' &\,\,~m' & m
\\
\,\,m\epsilon^2 &\,\,~~ m &\,\,~m &~~m_{\nu_s}
\end{array}\right) }~.
\end{array}  \!\!  ~~~~~
\label{matst}
\end{equation}

From (\ref{not}), taking $\kappa'\sim 10^{-3}$,
$\epsilon \simeq  0.2$, one has

\beq
m\simeq 1{\rm eV}~,~~~~m'\simeq 10^{-3}{\rm eV}~,~~~~
m_{\nu_s}\simeq 2\cdot 10^{-3}{\rm eV}~.
\label{val}
\eeq

From (\ref{matst}) and (\ref{val}),

$$
\Delta m_{31}^2\simeq m'^2\simeq 10^{-6}{\rm eV}^2~,
$$
\beq
{\cal A}(\nu_e\to \nu_{\mu , \tau })=\sin^2 2\theta_{e\tau }
\sim4\epsilon^4
\simeq 6\cdot 10^{-3}~,
\label{solmsw}
\eeq

$$
\Delta m_{\nu_s 2}^2\simeq 2mm_s
\simeq 3\cdot 10^{-3}{\rm eV}^2~,
$$
\beq
{\cal A}(\nu_{\mu }\to \nu_s)=1-
{\cal O}\left(\frac{m_{\nu_s}^2}{m^2}\right)~.
\label{atmst}
\eeq
We see that maximal $\nu_{\mu }-\nu_s$ mixing is realized by a proper
choice of the ${\cal U}(1)$ charges of the appropriate fields. The
sterile neutrino is kept light with the help of the 
${\cal U}(1)$ symmetry \cite{jo,422}. Having a lower value
for $m_{\nu_s }(\sim 10^{-3}$~eV), 
we can
still obtain the same value for $\Delta m_{\nu_s 2}^2 $ for $m\sim 3$~eV
(this value for $m$ in (\ref{val}) is obtained for $\epsilon \simeq
0.21$). 
In this case one of the active neutrinos  has mass
$m_{\nu_2 }\sim 3$eV, and therefore contributes roughly $15\%$ to
the critical energy density of the universe.
Models of structure formation with cold and hot dark matter
\cite{hdm} are in good agreement with the observations.

In summary, we have shown how different scenarios 
for large (even maximal) $\nu_{\mu}-\nu_x $ neutrino mixings
are realized 
in supersymmetric $SU(5)$ in order to accommodate the atmospheric
neutrino data. The solar neutrino puzzle can be explained by either
maximal angle vacuum oscillations (bi-maximal scenario), 
or through the small mixing angle MSW solution.

We now move on to the  problematic asymptotic mass relations
$m^{(0)}_{\mu}=m^{(0)}_s$ and $m^{(0)}_e=m^{(0)}_d$.
These arise from the coupling
$10\cdot\bar{5}\cdot\bar{H}$ which contains the mass generating terms
$qd^ch_d$ and $le^ch_d$.
In order to break this mass degeneracy (we would like to retain
$m^{(0)}_{\tau}=m^{(0)}_b$), let us introduce a pair of
`matter' superfields belonging to the $15+\overline{15}$ of $SU(5)$
where, under
$SU(3)_c \times SU(2)_L$,
\beq
15\ =\ (3,2)+(6,1)+(1,3)\ .
\label{15plet}
\eeq
It is clear that with $\Sigma 10\overline{15}$ type couplings,
only $q$ states from $10$ plets will mix with the
corresponding states in $15$. This
only affects the down and up quark mass matrices, but not  the charged
lepton Yukawa couplings. This mixing enables us to break the down
quark-charged lepton mass degeneracy. Let us note that this mechanism
differs from those suggested previously \cite{su52}.

With the following prescription of the ${\cal U}(1)$ charges of
$(15+\overline{15})_{1,2}$ and $\Sigma $ multiplets

$$
Q_{15_1}=q_2-6~,~~~Q_{15_2}=q_2-5~,~~~Q_{\Sigma }=0~,
$$
\beq
~~~~~Q_{\overline{15}_1}=1-q_2~,~~~~Q_{\overline{15}_2}=2-q_2~,
\label{charges15}
\eeq
the relevant couplings are

\begin{equation}
\begin{array}{cc}
 & {\begin{array}{ccc}
10_1&\,\,10_2&\,\,10_3~~~
\end{array}}\\ \vspace{2mm}
\begin{array}{c}
\overline{15}_1\\ \overline{15}_2

\end{array}\!\!\!\!\! &{\left(\begin{array}{ccc}
\,\, \epsilon^2~~&
\,\,  \epsilon~~ &\,\, 0
\\
\,\, \epsilon ~~ &\,\,1~~&\,\, 0~
\end{array}\right)\Sigma }~,
\end{array}  \!\!~
\begin{array}{cc}
 & {\begin{array}{cc}
15_1&\,\,
~15_2~~~~~
\end{array}}\\ \vspace{2mm}
\begin{array}{c}
\overline{15}_1 \\ \overline{15}_2

\end{array}\!\!\!\!\! &{\left(\begin{array}{ccc}
\,\, \epsilon^5~~
 &\,\,\epsilon^4
\\
\,\, \epsilon^4~~
&\,\,\epsilon^3
\end{array}\right)M_P.
}
\end{array}
\label{1510}
\end{equation}
Through (\ref{charges15}) choice and (\ref{1510}) couplings, $q_2$
is undetermined and therefore, charged fermion sector is consistent
with various oscillation scenarios discussed above.

Choosing a basis in which the couplings (\ref{lepdown}) and 
the mass matrix for $15$-plets (in (\ref{1510}))
are diagonal,
the mass matrix relevant for down quarks is given by 

\begin{equation}
\hat{M_d}=\begin{array}{cc}
 & {\begin{array}{cc}
~~~q_{10}~~  &\,\,~~q_{15}~~~~
\end{array}}\\ \vspace{2mm}
\begin{array}{c}
d^c \\ \bar q_{\overline{15}}
\end{array}\!\!\!\!\! &{\left(\begin{array}{cc}
\,\, \hat{Y}_e^Dh_d &\,\,~~0
\\
 \,\, \hat{C} &\,\,~~\hat{M}_{15}

\end{array}\right)}~~,
\end{array}   ~~
\label{down1}
\end{equation}
where
\begin{equation}
\begin{array}{ccc}
\hat{C}=~~ \\
\end{array}
\hspace{-6mm}\left(
\begin{array}{ccc}
c_{11}\epsilon^2& c_{12}\epsilon & c_{13}\epsilon^3 \\
c_{21}\epsilon & c_{22} &c_{23}\epsilon^2
 \end{array}
\right)M_P\epsilon_G  ~,~~
\begin{array}{ccc}
\hat{M}_{15}=~~ \\
\end{array}
\hspace{-6mm}\left(
\begin{array}{ccc}
\epsilon^5& 0  \\
0 & \epsilon^3
\end{array}
\right)M_P ~,
\label {AM}
\end{equation}
and
\begin{equation}
Y_e^D={\rm
Diag}\left(a_1\epsilon^5~,~a_2\epsilon^2~,~a_3\right)\epsilon^a~
\label{lepdiag}
\end{equation}
($\epsilon_G\equiv \langle \Sigma \rangle /M_P$).
For $c_{12}\stackrel{_<}{_\sim }1/5(\simeq \epsilon )$ 
and other couplings of
order unity, integration in (\ref{down1}) of the heavy states leads to the
down quark mass matrix:
\begin{equation}
\begin{array}{ccc}
 & {\begin{array}{ccc}
\hspace{-5mm}q_1'~~~~~ & \,\,q_2' ~~~ & \,\,q_3~

\end{array}}\\ \vspace{2mm}
\hat{m}_d= \begin{array}{c}
d^c_1 \\ d^c_2 \\d^c_3
 \end{array}\!\!\!\!\! &{\left(\begin{array}{ccc}
\,\,\lambda''\epsilon^5~~ &\,\,\lambda_2\epsilon^5~~ &
\,\,\epsilon^6
\\
\,\,\lambda_1\epsilon^3~~   &\,\,\lambda'\epsilon^2~~  &
\,\,\epsilon^4
 \\
\,\,0~~ &\,\,0~~ &\,\,a_3
\end{array}\right)\epsilon^ah_d }~,
\end{array}  \!\!  ~~~~~
\label{down}
\end{equation}
where $\lambda_{1,2}$, $\lambda'$ and $\lambda'' $ are some coupling
constants.  From (\ref{lepdiag}) and (\ref{down}) we have
the desired relation

\begin{equation}
\lambda_b=\lambda_{\tau }=a_3\epsilon^a~,
\label{btau}
\end{equation}
while the unwanted mass degeneracy between the down quarks and 
charged leptons of the two `light' families are avoided.
Note that the desired ratios
(see (\ref{lambdas})) between the
down quark Yukawa couplings still hold. Analyzing the  couplings
in (\ref{1510}),
one can verify  that the `light' left handed quarks $q'_{1,2}$ reside
in
$10_{1,2}$  and $15_{1,2}$ states (respectively), weighted in
comparable order. 
This means that the hierarchical structure of the couplings
(\ref{up}) will not be altered, and the magnitudes of the up 
type quark Yukawa
constants will still be given by (\ref{lambdas}).

Finally, let us note that by choosing basis (\ref{AM}), (\ref{lepdiag}),
and using (\ref{up}), (\ref{down})
the desired magnitudes of the CKM matrix elements (see (\ref{ckm}))
are obtained.
Furthermore, the unification of the three gauge coupling 
constants at $M_{GUT}$ is 
not affected, since all the additional fragments form complete $SU(5)$
multiplets, and decouple without significant splitting.

In conclusion, we have shown how a variety of neutrino oscillation
scenarios, that are consistent with the experimental data from
solar and atmospheric neutrino experiments, can be realized within a
$SU(5)$ scheme supplemented
by a ${\cal U}(1)$ flavor symmetry and some gauge singlet (right
handed) neutrinos.
We showed , in particular, how bi-maximal mixings can
be achieved without fine tuning the parameters. Two scenarios for
realizing the small mixing angle MSW solution, with one involving a
sterile state and giving possibility of existence of the neutrino
hot dark matter, are also displayed.  Note that in this scheme
neutrinoless double
$\beta $ decay and other flavor changing processes are strongly
suppressed.

Finally, consistent with these scenarios, a new
mechanism for resolving the
unacceptable mass relations involving the two light
families (of down quarks and charged leptons) is presented.
This was achieved by introducing the $15$-plets of $SU(5)$ \cite{ald}. 
The status of nucleon decay in this modified  scheme is essentially the
same as for minimal $SU(5)$.

\vspace{0.3cm}

This work was supported in part by  DOE under Grant No. DE-FG02-91ER40626
and by NATO, contract number CRG-970149.

\bibliographystyle{unsrt}

\end{document}